\documentclass[11pt,a4paper]{article}
\usepackage{jheppub}
\usepackage{amsmath}

\newcommand*\PD[1]{\doswap#1\relax}
\def\doswap#1,#2,#3\relax{\left( \frac{\partial #1}{\partial #2}\right)_{#3} }
\newcommand{\Mpl}{M_\mathrm{Pl}}
\newcommand{\rmax}{r_\mathrm{max}}
\newcommand{\amax}{a_\mathrm{max}}

\title{Holographic thermalization\\and Oppenheimer-Snyder collapse}

\author[a,b]{Olli Taanila}

\affiliation[a]{Nikhef, Science Park 105, 1098 XG, Amsterdam, The Netherlands}
\affiliation[b]{Helsinki Institute of Physics, P.O.Box 64, FI-00014 University of Helsinki, Finland}

\emailAdd{olli.taanila@iki.fi}

\abstract{The Oppenheimer-Snyder model, which describes the gravitational collapse of a ball of dust, can be used to model thermalization of strongly coupled systems using the AdS/CFT-duality. It can be used as an alternative to the previously widely used thin-shell model in the context of holographic thermalization. We solve the dynamics of the Oppenheimer-Snyder collapse in asymptotically AdS space-times and compute the two-point function of a boundary spectator field in the geodesic approximation. As a comparison we perform the same computation in the thin-shell model and discover oscillatory solutions of the thin-shell model for certain equations of state.}


\begin{document}

\begin{flushright}
NIKHEF-12\\HIP-2015-22/TH
\end{flushright}

\maketitle

\flushbottom

\section{Introduction}

Holographic methods have been well established in the recent years as a serious tool to understand the dynamics and thermalization of strongly coupled systems. The AdS/CFT duality \cite{Maldacena:1997re,Witten:1998qj,Gubser:1998bc} maps the complex process of thermalization of a strongly coupled $\mathcal{N}=4$ SYM field theory in four dimensions into the relatively simple process of gravitational collapse in 5D gravity. The use of the duality has spawned an abundance of interesting work attempting to gain insight to different aspects of thermalization (e.g.~see references 8-37 in \cite{Keranen:2015fqa}). Work that has attempted to describe states close to those in heavy ion collisions has been performed in e.g.~\cite{Chesler:2010bi,vanderSchee:2013pia,Casalderrey-Solana:2013sxa,Wu:2011yd,Chesler:2015wra}, in the sense that the initial state and the distribution of the energy-momentum in the boundary field theory is close to the one in heavy ion collisions. Although certain simple quantities, such as the aforementioned energy-momentum tensor, are easily computed in these numerical works, the computation of other observables can be very difficult due to the numerical nature of the work. For computing other observables analytical models of the thermalization process are thus favoured. The general theme of these analytical models is then that although their symmetries prohibit them describing the initial state of a heavy ion collision accurately, they are simple enough that the gravitational dynamics can be understood analytically and computations involving different observables can be performed in these backgrounds.

The simplest analytical model of gravitational collapse used in holography is the thin-shell model \cite{Danielsson:1999fa,Danielsson:1999zt}. It consists of an infinitely thin shell of matter with some equation of state collapsing into a black hole. A variation of this simple model is the Vaidya metric, where the shell is null (see \cite{Hubeny:2013dea} and references therein). The relative simplicity of the thin-shell model enables the computation of different quantities such as higher correlations functions of the stress tensor \cite{Lin:2008rw}, photon and dilepton production rates \cite{Baier:2012ax,Baier:2012tc}, the two-point function in the geodesic approximation \cite{Balasubramanian:2010ce,Balasubramanian:2011ur}, and different kinds of entropies \cite{Keranen:2014zoa,Keranen:2015fqa,Baron:2012fv,Baron:2013cya} just to name a few.

The initial state described by the thin-shell collapse however has a peculiar feature: Due to the translational symmetry of the setup, all the one-point functions in the boundary field theory have their equilibrium value throughout the evolution. This is in stark contrast with the initial state of heavy ion collisions, where the initial state is described by a very inhomogeneous distribution of matter. Thus the thin-shell model cannot describe perhaps the most important observable, the thermalization of the energy-momentum tensor. Thus it is clear that the thin-shell model cannot describe in detail heavy ion collisions or other processes encountered in the laboratory setting. Instead, we hope that by studying these analytical models, we can learn something qualitative about the dynamics of the strongly coupled system. When we compute observables, we hope that their features are generic for thermalization processes, and can be applied to other scenarios as well. 

It is however not clear which features of the predictions of the thin-shell model are generic and which very specific to that setup. For example, the thin-shell model has an energy-momentum tensor that is zero throughout in the bulk except for a delta function contribution of the shell, and, in fact, certain observables such as the photon and dilepton production rates hves been observed to have oscillatory features \cite{Baier:2012ax,Baier:2012tc} for which it is unclear whether they are generic signatures of thermalization or whether they are caused by the rather singular nature of the shell. For the forementioned reasons it is extremely important to employ different analytical models for holographic thermalization. Ideally, we could compute observables in different scenarios and thus find out which predictions seem generic patterns of thermalization and which features are specific to each model.

An alternative to the thin-shell model is the Oppenheimer-Snyder (OS) collapse. This describes the gravitational collapse of a homogeneous and isotropic ball of dust initially at rest. This model was first published already in 1939 \cite{Oppenheimer:1939ue} as a model of star formation in 3+1 dimensional asymptotically Minkowski space-time. In the context of AdS/CFT it was first mentioned in the appendix of \cite{Giddings:2001ii}, but aside from the equations of motion of the collapse it was not developed further there. The advantage of the Oppenheimer-Snyder collapse in the context of holography is that since it describes a ball of dust instead of an infinitely thin shell of matter, the energy-momentum tensor does not have a delta function, but rather is finite everywhere. Thus the resulting space-time is more regular at the edge of the ball than at the location of an infinitely thin shell. The model is still symmetric enough that for example the equations of motion of a spectator field in the bulk can solved at least for some cases.

It is the study of the Oppenheimer-Snyder collapse as a model of holographic thermalization that this work is devoted to. We will first introduce the model and solve its gravitational evolution. Then as a test observable, we compute the equal time two-point function in the geodesic approximation. To compare the OS setup with the thin-shell model, we perform the same computations in the thin-shell model. As a by-product of the computation, we discover previously unknown oscillatory solutions of the thin-shell model in global AdS.

The paper is organized as follows: In section \ref{sec:OS} we introduce the Oppenheimer-Snyder model in asymptotically AdS space-times, and discuss its dynamics. In section \ref{sec:ts} we review the thin-shell model and review the dynamics of the collapse. In section \ref{sec:correlator} we compute the equal time two-point correlator in the geodesic approximation in both of the models and compare the results. Finally in section \ref{sec:conclusions} we conclude and discuss possible developments.

\section{Asymptotically Anti-de Sitter Oppenheimer-Snyder collapse}

\label{sec:OS}

The Oppenheimer-Snyder collapse is conceptually the simplest model of gravitational collapse: A homogeneous and isotropic ball of pressureless dust undergoing gravitational collapse. Since we are looking for spherically symmetric solutions, we use global coordinates for AdS, where constant-$r$ hypersurfaces have explicit rotational symmetry.

Outside the ball the energy-momentum content is zero. The solution with zero $T_{\mu\nu}$, negative cosmological constant and rotational symmetry is given by the AdS-Schwarzschild solution:
\begin{equation}
\label{eq:outsidemetric}
ds^2 = -f(r) \, dt^2 + \frac{dr^2}{f(r)}+r^2\,d\Omega_n^2 \; , \quad f(r) = 1 + \Lambda\,r^2 -\frac{m}{r^{n-1}} \; ,
\end{equation}
where $d\Omega_n^2$ is the metric of a $n$-sphere and $\Lambda$ is the cosmological constant. In this coordinate system the edge of the sphere has some time dependent value of the radial coordinate $r_s$. Thus this coordinate patch is valid for $r> r_s$.

Inside the ball of dust the simplest coordinate frame to use is the rest frame of the dust. Since the dust is homogeneous and isotropic, we want a solution which has only time-dependent energy-momentum content, negative cosmological constant and rotational symmetry. This metric is of course the hyperbolic Friedmann-Robertson-Walker metric,
\begin{equation}
 ds^2 = -d\eta^2 + a(\eta)^2 \left[ \frac{d\rho^2}{1+\rho^2} + \rho^2 d\Omega_n^2\right]  \; .
\end{equation}
Imposing the Einstein equation with a negative cosmological constant and a pressureless fluid at rest and then using the energy-momentum continuity for the fluid, we arrive at the Friedmann equation
\begin{equation}
\label{eq:friedmann} \left(\frac{da}{d\eta}\right)^2 = 1 + \frac{2}{n(n+1)\Mpl^2}\epsilon_0\frac{a_0^{n+1}}{a^{n-1}}-\Lambda a^2 \; ,
\end{equation}
where $\epsilon_0$ is the energy density of the dust when the scale factor has the value $a_0$. Since this is in the dust's rest frame, and its total quantity is conserved, the edge of the ball sits at a constant value of the radial coordinate $\rho$, which we shall call $\rho_0$. Thus the validity of this coordinate patch is $0 < \rho < \rho_0$.

In order to glue these two coordinate patches together, we relate the two different patches together at the edge of the sphere and then require that the induced metric there is the same in both of the coordinate systems and that the extrinsic curvature is continuous across the edge of the sphere.
These conditions are the  Israel junction conditions \cite{Israel:1966rt} with zero energy-momentum content at the hypersurface joining the two patches. Although this might seem like heavy machinery for the problem at hand, we use it since it is applicable later on when we consider the thin-shell case as a comparison to the OS model.

The coordinates describing the edge of the sphere are chosen to be the angular coordinates, which are identified to be the same inside and outside, and the proper time at the edge, $\tau$. From the induced metric we can see that we can identify the time coordinate $\eta$ at the edge of the sphere with the proper time at the edge, i.e.~$\eta_s = \tau$. From that it follows that $r_s(\tau) = \rho_0 a(\tau)$. Additionally the time and radial coordinates $t_s$ and $r_s$ are related by \[-1 = -f(r_s)\,\dot{t}_s^2 + \dot{r}_s^2/f(r_s) \; .\]
Next we use the Israel junction conditions to relate the curvature on both sides. The junction conditions state that the difference of the curvature\footnote{Note that there is an ambiguity in the literature related to the sign of $K_{ij}$. For example \cite{Poisson} defines $K_{ij}$ with an opposite sign. In this case the sign ambguity does not matter, but can cause confusion when the energy-momentum of the joining hypersurface is non-zero.}
\begin{equation}
K_{ij} = n_\alpha \left( \frac{\partial^2 X^\alpha}{\partial\xi^i\,\partial\xi^j}+\Gamma^\alpha_{\beta\gamma}\frac{\partial X^\beta}{\partial\xi^i}\frac{\partial X^\gamma}{\partial\xi^j} \right)
\end{equation}
across the joining hypersurface is related to the energy-momentum content at the hypersurface, which in this case is zero. Here $X^\alpha$ are the $n+2$-dimensional coordinates of the $n+1$-dimensional hypersurface parametrized by the $n+1$ coordinates called $\xi^i$, and $n^\alpha$ is the normal vector of the hypersurface. $\gamma_{ij}$ is the induced metric on the surface. Since the energy-momentum content is zero, the extrinsic curvatures inside and outisde are related by
\begin{equation}
\left(K_{ij} - K\,\gamma_{ij}\right)_\mathrm{inside} = \left(K_{ij} - K\,\gamma_{ij}\right)_\mathrm{outside} \; .
\end{equation}
A straightforward computation gives the non-zero components of the above expression in the inside coordinate system to be
\begin{align}
K_{\tau\tau} -K\,\gamma_{\tau\tau} &= -n \frac{\sqrt{1+\rho_0^2}}{\rho_0 a(\tau)}\\
K_{\phi_n \phi_n} -K\,\gamma_{\phi_n \phi_n} & = g_{\phi_n \phi_n} (n-1) \frac{\sqrt{1+\rho_0^2}}{\rho_0 a(\tau)} \; ,
\end{align}
while in the outside coordinate system the non-zero components are given by
\begin{align}
\label{eq:Koutside1} K_{\tau\tau} -K\,\gamma_{\tau\tau} &= -n\frac{\sqrt{f(r_s)+\dot{r}_s^2}}{r_s}\\
\label{eq:Koutside2} K_{\phi_n \phi_n} -K\,\gamma_{\phi_n \phi_n} & = g_{\phi_n \phi_n} \frac{1}{r_s\,\dot{r}_s}\frac{d}{d\tau} \left( r_s^{n-1}\sqrt{f(r_s)+\dot{r}_s^2}\right)\; .
\end{align}
Equating the expressions of the inside and outside coordinate system, we get the simple identity
\begin{equation}
\label{eq:OSeom}
f(r_s) + \dot{r}_s^2 = 1 + \rho_0^2 \; .
\end{equation}
Since $\rho_0$ is a constant, this is an equation of motion for the radial position of the shell in the outside coordinate system, $r_s$. Substituting $r_s = \rho_0 a(\tau)$ into this equation, we get an equation of motion for the scale factor $a$. If one substitutes the explicit functional form of $f$, this can be seen to be exactly the Friedmann equation, already given in equation (\ref{eq:friedmann}).

In principle one can proceed with the above derivation with also non-zero pressure of the fluid the sphere is composed of, however, one will then see that a solution satisfying both equations (\ref{eq:friedmann}) and (\ref{eq:OSeom}) can only be found if $p=0$. This is not surprising, since if the pressure of the fluid would be non-zero, there would be a pressure gradient at the edge of the sphere and the fluid would no longer stay in rest in the inside coordinate system.

\begin{figure}
\centering
\includegraphics[scale=0.32]{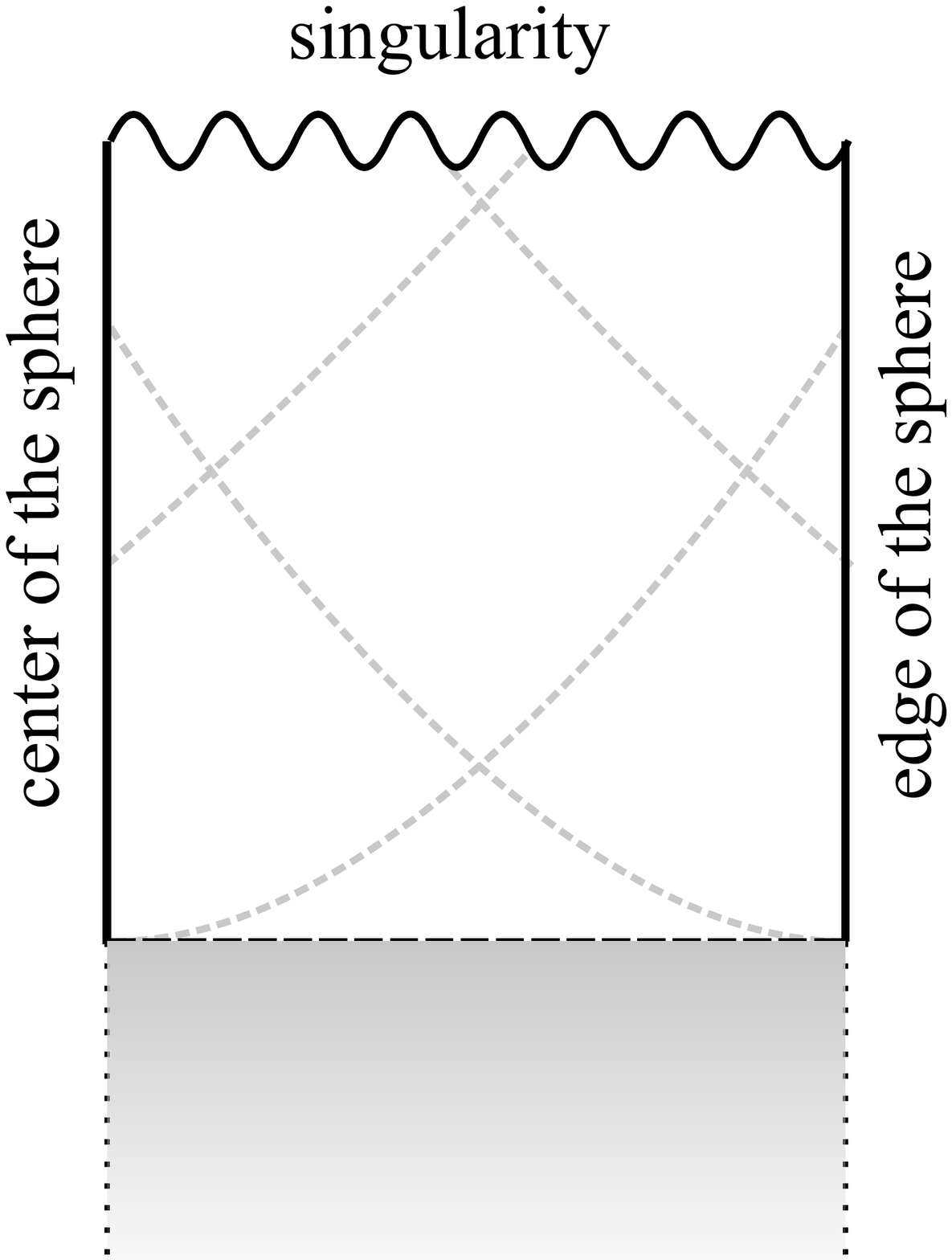}
\caption{A schematic of the coordinate system $(a,\rho)$ inside the sphere. The grey dashed lines are null geodesics. The grey region is the lower half of the manifold, where the sphere is expanding, however, the $a$-coordinate covers only the upper (or lower, depending on the sign choice) region. Note that there are null rays emanating from the center of the sphere which hit the singularity before encountering the edge of the sphere.}
\label{fig:inside}
\end{figure}

\subsection{The motion of the Oppenheimer-Snyder sphere}

Next we analyse the motion and the causal structure of the Oppenheimer-Snyder collapse. To do that, we use inside the scale factor as the time coordinate, that is, we make the coordinate substitution $da = \frac{da}{d\eta}\, d\eta$, so that the metric becomes
\begin{equation}
\label{eq:insidemetric} ds^2 = -\frac{da^2}{h(a)} +\frac{a^2}{1+\rho^2} d\rho^2 +a^2\,\rho^2\,d\Omega_n^2 \; ,
\end{equation}
where the metric function $h(a) = \left(\frac{da}{d\eta}\right)^2$ comes directly from the Friedmann equation (\ref{eq:friedmann}). It is convenient to relate this function to $f$ via
\begin{equation}
h(a) = 1+\frac{1-f(r_s)}{\rho_0^2} \; .
\end{equation}

The time evolution of the diameter of the sphere can be found from equation (\ref{eq:OSeom}). The first observation is that since $f(r)$ grows without bound, for any $\rho_0$ there is always a turning point $\rmax$, where $\dot{r}_s =0$. Furthermore, since
\begin{equation}
\frac{d^2r_s}{d\tau^2} = \frac{1}{2}\frac{d}{dr_s} \dot{r}_s^2 = -f'(r_s) \; ,
\end{equation}
the acceleration at the turning point is always negative, and at the turning point the sphere starts to collapse. This means that there is also a maximum value of the scale factor, $\amax$, which is the value of the scale factor at the turning point. The space-time also has a space-like singularity at $a=0$, which is the ``big crunch'' of the space-time.

Due to the time reversal symmetry of the setup, in principle the evolution of the sphere is such that it emerges from a singularity and expands until it reaches $\rmax$, after which it collapses back to a singularity. Thus the coordinate system $(a,\rho)$ covers only one half of the manifold. We consider only the upper half of the space-time which describes the collapse of the shell starting form the turning point, with the view that the system is initially prepared to the state described by the sphere at the turning point.

It is also important to note that while there is some value of the ``time coordinate'' $a=a_h$ when the horizon is formed, there is no qualitative change inside. In fact, although the value of $r_h$ and thus $a_h$ can be changed by tuning the value of $\rho_0$, the Friedmann equation is indifferent to the choice of that parameter.

\section{The thin-shell model}

\label{sec:ts}

The thin-shell model has been discussed widely in the literature \cite{Baron:2012fv,Lin:2008rw,Baier:2012ax,Baier:2012tc,Keranen:2014zoa,Keranen:2015fqa} in the context of holographic thermalization. Another variation on the same theme is the Vaidya metric, where the collapsing shell is null (see e.g.~\cite{Hubeny:2013dea}). However, in most of these works the collapse has been studied in the Poincar\'{e} patch of the AdS space. In this coordinate system the shell, which is located at constant value of the $r$-coordinate, is actually an infinite flat sheet with translational symmetry. To enable a comparison with the Oppenheimer-Snyder collapse, we need to use the global coordinate system of the AdS-space, where a constant-$r$-hypersurface is an actual spherical shell with rotational symmetry. It turns out that the spherically symmetric case has interesting behaviour not present in the previously studied case in the Poincar\'{e} patch. The notation follows closely that of \cite{Keranen:2015fqa}, so we suggest the reader to refer to that one for more details. 

The metric of the thin-shell model in global AdS is given by
\begin{equation}
\label{eq:thinshellmetric} ds^2 = -f(r)\,dt^2 +\frac{dr^2}{f(r)}+r^2\,d\Omega_n^2
\end{equation}
where the metric function $f$ is now defined by
\begin{equation}
\label{eq:thinshellfpm} f =
\begin{cases}
f_- = 1+\Lambda r^2\,, & r<r_s\\
f_+ = 1+\Lambda r^2 - \frac{m}{r^{n-1}}\,, & r>r_s \end{cases}
\end{equation}
and the location of the shell $r_s$ is time dependent. The radial coordinate $r$ and the angular coordinates are continuous at the shell, but the time coordinate is discontinuous, and thus there are two different time coordinates, $t_-$ and $t_+$, the time inside and outside of the shell respectively. To parametrize the shell, we use the angular coordinates and the proper time of the shell, $\tau$, and use an overdot to denote a derivative with respect to $\tau$.

Unlike in the OS collapse, now both the inside and outside solutions are vacuum solutions, and all the energy-momentum content is located at the shell. This means that the Israel junction conditions are now
\begin{equation}
\label{eq:thinshellisrael} \left(K_{ij}- K \,\gamma_{ij} \right)_\mathrm{inside} - \left(K_{ij}- K \,\gamma_{ij} \right)_\mathrm{outside} = -8\pi g_5 S_{ij} \; ,
\end{equation}
where $S_{ij}$ is now the energy-momentum tensor of the shell in the $\xi$-coordinate system. We use the ideal fluid form,
\begin{equation}
S^{ij} = (\rho+p)u^i u^j + p\gamma^{ij} \;, \quad p = c\rho \; ,
\end{equation}
where $u^i = \delta^i_0$, since the fluid is at rest in the $\xi$-coordinate system. The expressions from equations (\ref{eq:Koutside1}) and (\ref{eq:Koutside2}) now apply both inside and outside, $f_-$ and $f_+$ substituted for $f$ respectively. Plugging these into the junction conditions of equation (\ref{eq:thinshellisrael}), we get two independent equations:
\begin{align}
-\frac{n}{r_s} \left( \sqrt{f_-(r_s)+\dot{r}_s^2}-\sqrt{f_+(r_s)+\dot{r}_s^2}\right) & = -8\pi g_5 \rho\\
g_{\phi_n\phi_n}\frac{1}{\dot{r}_s\,r_s}\frac{d}{d\tau} \left[ r_s^{n-1}\left( \sqrt{f_-(r_s)+\dot{r}_s^2}-\sqrt{f_+(r_s)+\dot{r}_s^2}\right) \right] & = -8\pi g_5 g_{\phi_n\phi_n} p
\end{align}
These equations hold for arbitrary relation between $\rho$ and $p$. We however assume a constant equation of state $c$, and thus the above equations are solved by
\begin{equation}
\label{eq:thinshelleom} \sqrt{f_-+\dot{r}_s^2}-\sqrt{f_++\dot{r}_s^2} = M\,r_s^{1-n(1+c)}\;,
\end{equation}
where $M$ is now an integration constant characterizing the energy of the shell. From this we can also solve the equation of motion for the shell,
\begin{equation}
\label{eq:rsdot2} \dot{r}_s^2 = \frac{M^2}{4}r_s^{2(1-n(1+c))} - \frac{f_-+f_+}{2}+\frac{r_s^{2(n(1+c)-1)}(f_--f_+)^2}{4M^2}\;.
\end{equation}
The integration constant $M$ encodes information about the initial conditions of the shell's motion. Since we want to compare with the motion in the Oppenheimer-Snyder collapse, we assume the shell starts to fall from a turning point $r_0$ at $\tau=0$, where $\dot{r}_s = 0$. We can evaluate equation (\ref{eq:thinshelleom}) at the turning point, and thus we get
\begin{equation}
M = \frac{\sqrt{f_-(r_0)}-\sqrt{f_+(r_0)}}{r_0^{1-n(1+c)}}\; .
\end{equation}
This expression with equation (\ref{eq:rsdot2}) determines the motion of the thin shell.

\subsection{Trajectories of the thin shell}

Next we wish to inspect the possible trajectories of the shell.  Instead of solving the position of the shell $r_s$ explicitly as a function of the coordinate time $t_s$, it is sufficient to look at the expression $\dot{r}_s^2$ in equation (\ref{eq:rsdot2}). First of all, since $\dot{r}_s^2$ is a square of the physical velocity, it has to be positive. Additionally, since $\ddot{r}_s = \frac{1}{2}\frac{d}{dr_s}\dot{r}_s^2$, the same expression also encodes information of the acceleration for the trajectory. Thus by computing the acceleration at the turning point, 
\begin{equation}
\left.\frac{d^2r_s}{d\tau^2}\right|_{r_s=r_0}= - \left[ r_0 \Lambda + \frac{m}{r_0^n}\frac{n-1}{4} + \frac{n(1+c)-1}{4}\frac{M^2}{r_0^{n(1+c)}} - \frac{n\,c\,m^2\,r_0^{2cn-1}}{4\,M^2} \right] \; ,
\end{equation}
we can determine towards which direction the shell starts to move. As long as the expression in the square brackets is positive, the shell starts to fall at the turning point. However, if
\begin{equation}
 \frac{n\,c\,m^2\,r_0^{2cn-1}}{M^2} \; > \; 4\, r_0 \Lambda + (n-1)\frac{m}{r_0^n} + \left[n(1+c)-1\right]\frac{M^2}{r_0^{n(1+c)}} \; ,
\end{equation}
instead of collapsing, the shell starts to expand. Thus all trajectories can be divided to two categories: The ones where the shell starts to collapse, and the ones where the shell starts to expand at its turning point $r_0$.

\begin{figure}
\centering
\includegraphics[scale=0.6]{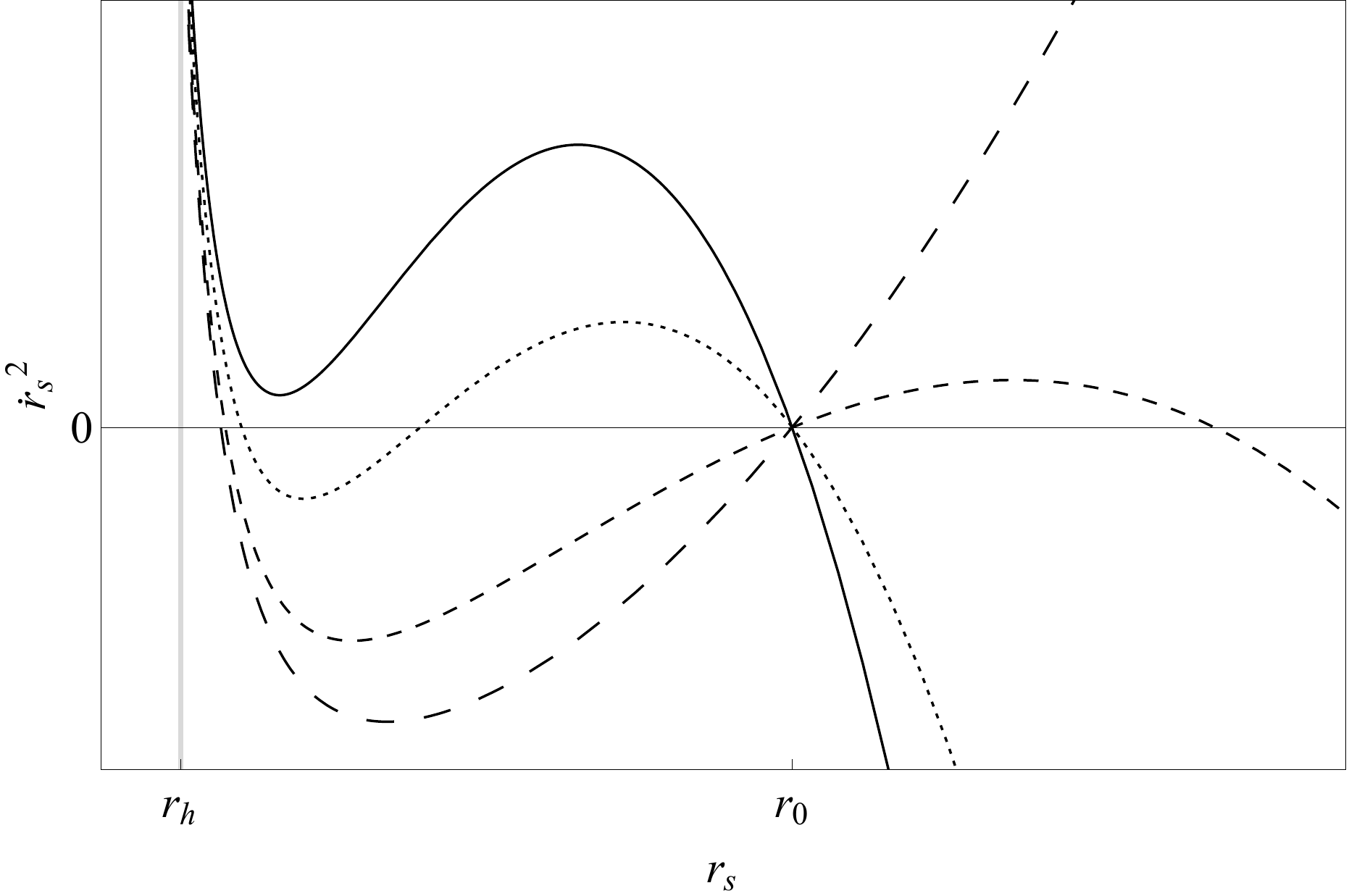}
\caption{The speed squared of the thin shell for different equations of state. Solid line $c=0.328$, the dotted line $c=0.33$, the small dashed line $c=0.3323$ and the large dashed line $c=0.3334$. The divergence of the velocity at $r_h$ is a sign of a coordinate singularity at the horizon. The other parameters have been chosen to be $m=5$, $\Lambda=1$, $r_0=10$. }
\label{fig:rsdotfig}
\end{figure}

Next question then is, what happens to the shell after its initial turning point? Is there another turning point, or will the shell always either collapse to a black hole or expand all the way to the boundary? The EOM cannot be solved analytically for turning points for general values of $n$ and $c$. When investigating the solutions numerically, it turns out that both kinds of behaviour are present. Thus the solutions can be divided into two additional categories: The ones which have only a single turning point, and the ones which have more. Thus we have four general categories of solutions, illustrated in figure \ref{fig:rsdotfig}:
\begin{enumerate}
 \item Shell collapses to a black hole (solid line)
 \item Shell starts to collapse, but has a turning point at smaller radius (dotted line)
 \item Shell starts to expand, but has a turning point at a greater radius (small dashed line)
 \item Shell expands all the way to the boundary (large dashed line)
\end{enumerate}
In case 1, the expression in equation (\ref{eq:rsdot2}) has only a single root. For case 2, the expression has three roots, however, since the initial condition is given at $r_0$, the smallest root is never reached, and the shell ends up oscillating between the $r_0$ and the middle root. For case 3 the expression has again three roots, and now the shell oscillates between $r_0$ and the largest root. For case 4, the the expression has two roots, but the smaller one is never reached, and instead the shell starts and continues to expand from $r_0$ towards infinity.

The different trajectories in figure \ref{fig:rsdotfig} differ only by equation of state. In the case of $c>-1$ and $n=3$, one can look at the powers of the different terms in equation (\ref{eq:rsdot2}): At small $r_s$ the first term will dominate no matter what the value of $c$. However, at large $r_s$, if $c<\frac{1}{n}$ the middle term will dominate and the expression will turn negative at sufficiently large $r_s$, and if $c>\frac{1}{n}$ then the last term will dominate and the expression will be positive at large $r_s$. Thus one critical value of $c$ is $\frac{1}{n}$, which separates cases 3 and 4. The critical values of $c$ separating cases 1 and 2, and 2 and 3, respectively, are however dependent on the other choices of parameters and initial conditions, and cannot be solved analytically. Thus the exact numerical values of the choices of $c$ in figure \ref{fig:rsdotfig} are not important, but instead it is sufficient to note that if one denotes $c_i$ as the equation of state of the different cases, then $c_1<c_2<c_3<c_4$, $c_3<\frac{1}{n}$ and $c_4>\frac{1}{n}$.

The trajectories of the shell encode three qualitatvely different effects: Geodesic motion in AdS, gravitational pull of the shell, and the pressure of the shell. The non-linear interplay of these different effects is the cause of this rich structure for the possible trajectories. The collapsing shell (case 1) and the eternally expanding shell (case 4) are the most familiar solutions, where as the eternally oscillating shell (cases 2 and 3) are more peculiar. In the context of holography it would be tempting to try to interpret these solutions as non-thermalizing states in the boundary field theory. However, when $c\neq0$, the shell is made out of matter that has pressure only in some spatial directions, and not in others. The classical analogy of this would be a sheet of rubber or a network of springs. In order for this shell to have a holographic interpretation in the boundary field theory, there must be a bulk degree of freedom from which the shell is constructed. Whether this is possible for the non-zero values of $c$ is an unanswered question. One familiar case where there is pressure along the shell is an instanton-like solution of e.g.~a scalar field in the bulk, but for these solutions the equation of state parameter is usually $c \simeq -1$.

As a final note, we must note that for the equation of state $c=0$, which one would imagine to be easiest to construct from the degrees of freedom available in the holographic description, the shell always collapses. Due to the lack of pressure, this is also the configuration we will use when making comparisons with the Oppenheimer-Snyder collapse in the following sections.

\section{The equal time two-point correlator}

\label{sec:correlator}

To probe the thermalization described by the Oppenheimer-Snyder collapse and the thin-shell collapse, we compute the equal time two-point correlator of an operator with a large conformal weight using the geodesic approximation. The geodesic approximation (for details and derivation see e.g.~\cite{Banks:1998dd,Balasubramanian:1999zv}) is essentially the saddle point approximation in the limit of large scaling dimension of the operator whose correlator we are computing. The description for the correlator is then given by
\begin{equation}
\label{eq:dictionary} \left\langle \mathcal{O}(\mathbf{x}) \mathcal{O}(\mathbf{x'})\right\rangle \simeq \epsilon^{-2\Delta}e^{-\Delta\sigma(\mathbf{x},\mathbf{x'})}
\end{equation}
where $\Delta$ is the conformal weight of the bulk particle or equivalently the scaling dimension of the dual operator, $\sigma(\mathbf{x},\mathbf{x'})$ is the length of the geodesic connecting the two boundary points $\mathbf{x}$ and $\mathbf{x'}$, and $\epsilon$ is a cutoff parameter regularizing the length of the geodesic as it approaches the boundary. From this we see that the computation of the equal time two-point correlator is reduced to computing the geodesic length between two points residing in the boundary at equal times. Due to the spherical symmetry of the setup, the geodesic length, and thus the correlator, will only depend on the time coordinate of the points and one angle separating the points in the $n$-sphere which forms the boundary. This angle we shall call $\varphi$.

The cutoff $\epsilon$ corresponds to some (large) finite radius $R$ to which the geodesics are computed. Since the divergence of the geodesic length when taking $R \to \infty$ is present even in the case of empty AdS, we define the renormalized geodesic length by
\begin{equation}
\tilde{\sigma}(t,\varphi) \equiv \sigma(t,\varphi) - \frac{2}{\sqrt{\Lambda}}\ln 2\sqrt{\Lambda} R \; ,
\end{equation}
where the last term on the RHS is the divergent part of the length of a spacelike geodesic in empty AdS. The renormalized $\tilde{\sigma}$ is thus finite in the limit $R\to\infty$. The renormalization scheme here is arbitrary, and in fact $\tilde{\sigma}$ will depend on the choice of the scheme. However, we will only compare the lengths of geodesics with matching boundary endpoints, and the choice of the renormalization scheme does not affect the \emph{difference} of the lengths of these geodesics. Similar scheme was used in e.g.~\cite{Hubeny:2013dea}.

%The normalization of $\tilde{\sigma}$ is somewhat arbitrary, since we could add or substract finite terms to the length, but since we are mainly comparing the geodesics in different geometries, the overall normalization does not matter as long as it is kept fixed in all the computations.

The derivation of the matching conditions of the geodesics in this section follows the same recipe and computational machinery described in \cite{Keranen:2015fqa}, appendices of which the reader is referred to for more detailed derivations.

\subsection{Calculational strategy}

We want to know the length of the shortest space-like geodesic connecting two boundary points with the time coordinate $t$ and the angular separation $\varphi$. Although a simple statement, this quantity is not trivial to compute, since if we start from the boundary points we do not know to which direction the geodesic should start to propagate from the boundary -- that behaviour encodes the bulk geometry of the space-time. Instead, we approach the problem from a different point of view: A geodesic that connects the two points with the same time coordinate must have a turning point somewhere in the bulk, where both the radial and temporal velocities\footnote{Here we use the term \emph{velocity} to denote a derivative of the coordinates of the geodesic with respect to the proper length of the geodesic. Although these quantities are not velocities in the physical sense, they appear in the geodesic equations in the same fashion as actual velocities would for time-like geodesics.} are zero. Thus we can start from any point in the bulk and look for geodesics which have a turning point at that point, and then propagate these geodesics all the way boundary. This means that for each set of initial conditions (in the Oppenheimer-Snyder case $(\bar{a},\bar{\rho})$ and in the thin shell case $(\bar{t}_-,\bar{r})$) we can compute the time $t$ when the geodesic hits the boundary, the angular separation $\varphi$, and the length of the geodesic $\sigma$. If we then scan through all possible values of the initial turning point, we should recover a geodesic length for any possible combination of $\varphi$ and $t$.

It turns out that one needs to consider also geodesics which penetrate the edge of the sphere or the thin shell \emph{after} the horizon has formed, i.e.~geodesics which cross the coordinate singularity at $r=r_h$ in the outside coordinate patch. Although the coordinate system has a singularity at that point, with careful regularization all physical quantities will stay finite and the computation can be carried out in this coordinate system. A more robust approach is to use the Eddington-Finkelstein coordinate system outside, details of which are give in Appendix \ref{sec:eddingtonfinkelstein}.

%It turns out that it is necessary to consider also geodesics that penetrate the shell or the edge of the sphere \emph{after} the horizon has formed, and thus traverse the black hole event horizon. To follow their evolution in the outside coordinate patch, SOMETHIGN HERE

%For a more transparent description, it is necessary to use the Eddington-Finkelstein coordinate system, details of which are given in appendix \ref{sec:eddingtonfinkelstein}.

\subsection{Geodesics in the Oppenheimer-Snyder collapse}

To calculate the length of geodesics connecting two equal time boundary points in the Oppenheimer-Snyder geometry, we need to solve the set of geodesic equations in the two coordinate systems defined by equations (\ref{eq:outsidemetric}) and (\ref{eq:insidemetric}), and then match the solutions at the edge of the sphere of dust.

The geodesic equations in the inside coordinate system are given by
\begin{align}
\label{eq:OSingeodesic1} -\frac{h'}{2\,h} \dot{a}^2 + a\,\rho^2\,h\,\dot{\phi}^2 + \frac{a \, h \, \dot{\rho}^2}{1+\rho^2} + \ddot{a} & \;=\; 0 \\
\label{eq:OSingeodesic2} -\rho\,(1+\rho^2)\,\dot{\phi}^2+\frac{2\,\dot{a}\,\dot{\rho}}{a}-\frac{\rho\,\dot{\rho}^2}{1+\rho^2} \ddot{\rho} & \;=\;0\\
\label{eq:OSingeodesic3} \frac{2\,\dot{a}\,\dot{\phi}}{a} + \frac{2\,\dot{\rho}\,\dot{\phi}}{\rho} + \ddot{\phi}& \; =\; 0 \; ,
\end{align}
where $'$ denotes a derivative w.r.t.~$a$ and $\dot{\;}$ a derivative w.r.t.~the affine parameter of the geodesic, which we choose to be the geodesic length $\sigma$.

The geodesic equations (\ref{eq:OSingeodesic1}-\ref{eq:OSingeodesic3}) can be integrated to give
\begin{align}
\label{eq:OSingeoint1} \dot{\phi} & \; = \; \frac{\bar{L}}{a^2\,\rho^2}\\
\label{eq:OSingeoint2} \dot{a}^2 & \; = \; h(a) \left[ \frac{\bar{E}}{a^2}-1\right]\\
\label{eq:OSingeoint3} \dot{\rho}^2 &\;=\; \frac{1+\rho^2}{a^4}\left[ \bar{E}-\frac{\bar{L}^2}{\rho^2}\right] \; ,
\end{align}
where $\bar{E}$ and $\bar{L}$ are integration constants, i.e.~constants of motion. Their names are chosen to remind the reader of energy and angular momentum, since they are the analogues of these quantities. The values of the constants of motion are determined from the requirement that the geodesic has a turning point at $(\bar{a},\bar{\rho})$:
\begin{align}
\dot{a}^2(a=\bar{a}) = 0 \qquad & \Rightarrow \qquad \bar{E} = \bar{a}^2\\
\dot{\rho}^2(\rho = \bar{\rho}) = 0 \qquad &\Rightarrow \qquad \bar{L} = \pm \bar{a}\bar{\rho}
\end{align}

\begin{figure}
\centering
\includegraphics[scale=0.9]{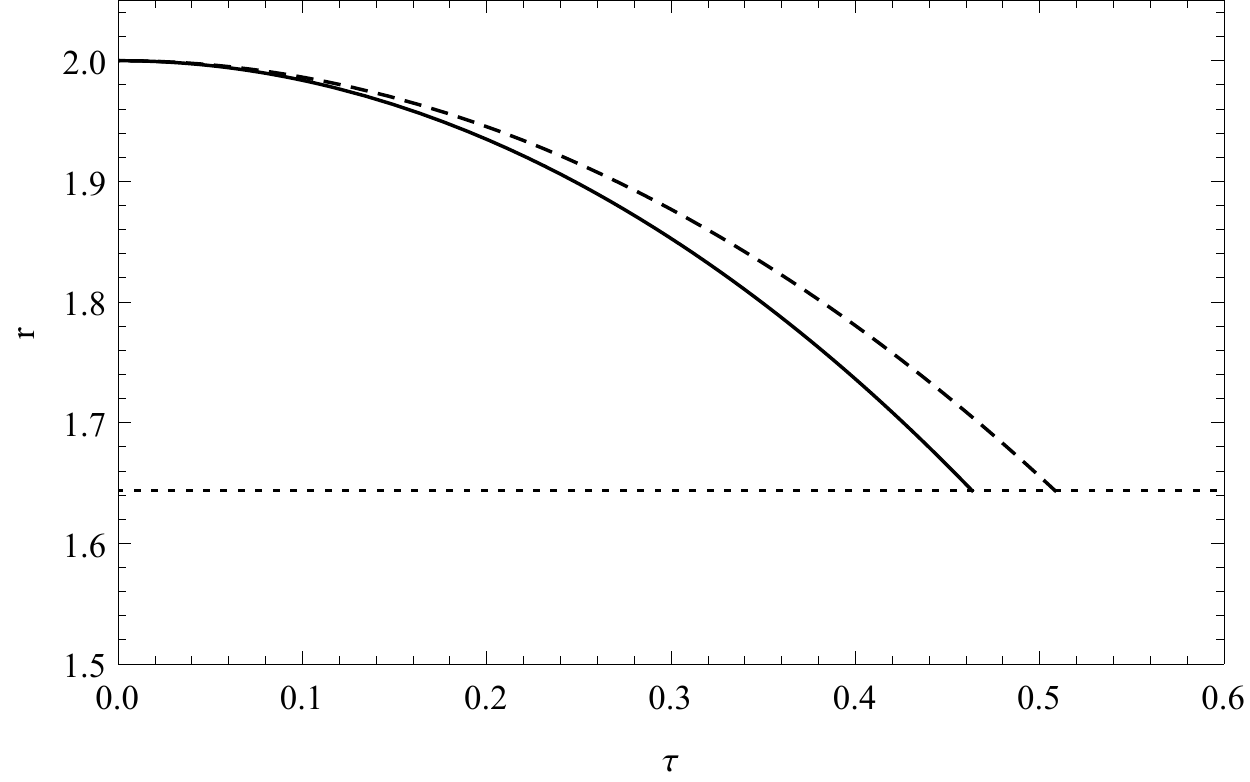}
\caption{Comparison between the trajectories of the pressureless ($c=0$) thin shell and the OS collapse with equal initial radius $r_0$ as a function of the proper time of the shell. Solid line is the position of the edge of the dust sphere of the OS collapse while the dashed line is the location of the thin shell. The dotted line is the location of the horizon. The difference between the two trajectories gets even smaller if $r_h$ is decreased or $r_0$ increased. Here the other parameters $\Lambda =1$, $m=10$, $n=3$ and $r_0=2$ have been chosen such that the difference is big.}
\label{fig:trajectories}
\end{figure}

Now that we know how the geodesic propagates from the turning point to the edge of the sphere, we need to solve how the geodesic propagates outside the sphere, and how the geodesic solution inside is matched to a solution outside. The geodesic equations outside are given by
\begin{align}
\frac{f'\,\dot{r}\,\dot{t}}{f} + \ddot{t} & = 0\\
-\frac{f'}{2\,f}\dot{r}^2 + \frac{1}{2}f\,f'\,\dot{t}^2 -r\,f\,\dot{\phi}^2 + \ddot{r} & = 0\\
\frac{2\,\dot{r}\,\dot{\phi}}{r}+\ddot{\phi} & = 0 \; .
\end{align}
These can be integrated to give
\begin{align}
\label{eq:outgeoint1} f\,\dot{t} &=E\\
\label{eq:outgeoint2} r^2\dot{\phi} & = L\\
\label{eq:outgeoint3} \dot{r}^2 & = f\left( 1-\frac{L^2}{r^2}\right)+E^2 \; .
\end{align}
To relate the constants of motion outside to the trajectory of the geodesic inside, we construct a coordinate system $(\tau,\lambda)$ which is continuous at the edge of the sphere and its close surroundings and use that to transform the geodesic derivatives in the inside coordinate system to those in the outside coordinate system. This computation is described in more detail in Appendix \ref{sec:continuous}, but its result is given by
\begin{align}
\label{eq:matchingOS1} E &= \sqrt{\frac{\bar{a}^2}{a^2}-1}\sqrt{1+\rho_0^2} + \frac{\bar{a}}{a}\sqrt{1-\frac{\bar{\rho}^2}{\rho^2}}\dot{r}_s\\
\label{eq:matchingOS2} L &= \bar{L} = \pm \bar{a}\bar{\rho} \; .
\end{align}
The procedure to compute the geodesic length for all times and angles is then as follows:
Given the initial conditions $\bar{a}$ and $\bar{\rho}$,
integrate equations (\ref{eq:OSingeoint1} - \ref{eq:OSingeoint3}) until the geodesic reaches the edge of the sphere.
Then use the matching conditions (\ref{eq:matchingOS1}) and (\ref{eq:matchingOS2}) to get the constants of motion outside.
With the constants of motion, integrate equations (\ref{eq:outgeoint1} - \ref{eq:outgeoint3}) until the geodesic reaches the boundary, which is represented by some (large but finite) cutoff in the radial coordinate, $R$.
Repeat the computation for all possible values of the initial conditions $\bar{a}$ and $\bar{\rho}$. One must discard those cases where either the geodesic does not exit the sphere before $a=0$, or the geodesic does exit the sphere, but turns around and falls back to the sphere and thus never reaches the boundary.

\subsection{Computation in the thin-shell model}

The computation in the thin-shell case is very similar to the Oppenheimer-Snyder case. Since the metric in equations (\ref{eq:thinshellmetric}) and (\ref{eq:thinshellfpm}) is the same as the metric of equation (\ref{eq:outsidemetric}), the geodesic equations both inside and outside are given by the equations (\ref{eq:outgeoint1} - \ref{eq:outgeoint3}). We shoot the geodesic from its turning point $(\bar{t}_-,\bar{r}_-)$. From the requirement that $\dot{r}=0$ and $\dot{t}=0$ at the turning point, we are able to derive that the constants of motion inside are given by
\begin{equation}
E_- = 0 \qquad \text{and} \qquad L_-=\pm \bar{r} \;.
\end{equation}
We can then integrate the geodesic equations (\ref{eq:outgeoint1} - \ref{eq:outgeoint3}) until the shell, $r = r_s$, at which point we need to switch to the geodesic equations in the outside coordinate patch. These are of course given by equations (\ref{eq:outgeoint1} - \ref{eq:outgeoint3}) as well, but now with $f=f_+$ and with different constants of motion. These constants can be derived (for more detail, see e.g. the Appendix B of \cite{Keranen:2015fqa}) to be
\begin{align}
\label{eq:matchingthinshell1} E_+ & = \sqrt{1-\frac{\bar{r}^2}{r_s^2}}\frac{\dot{r}_s}{\sqrt{f_-}}\left( \sqrt{f_-+\dot{r}_s^2}-\sqrt{f_++\dot{r}_s^2} \,\right)\\
\label{eq:matchingthinshell2} L_+ & = \bar{r} \; .
\end{align}
The procedure to compute the geodesic length is after this point very similar to the Oppenheimer-Snyder case: Given the initial radial coordinate $\bar{r}$, integrate the equations for some value of $r_s$. This corresponds to a choice of initial time. Then use the matching conditions (\ref{eq:matchingthinshell1}-\ref{eq:matchingthinshell2}) to compute the constants of motion outside, after which integrate the geodesic equations (\ref{eq:outgeoint1}-\ref{eq:outgeoint3}) until the geodesic reaches the 'boundary', i.e.~the radial cutoff $R$. Then rinse and repeat for all values of $\bar{r}$ and $r_s$. Again, one must discard geodesics which never reach the boundary and also take into account geodesics that exit the shell after the horizon has formed.

\subsection{Results}

We wish to compare the thermalization of Oppenheimer-Snyder to that of a thin-shell collapse. To make the comparison of the two qualitatively different models meaningful, we make the mass of the final black hole and the initial radius of the sphere and shell to be the same for both models. This corresponds to setting the final temperature of the state of the boundary field theory as well as the angular scale of the initial state to be the same for both states. Also, since the sphere of dust is pressureless by construction, we also choose the shell to be pressureless, i.e.~$c=0$. The differences in the correlator for the two different models are then due to two distinct sources: different speed of the gravitational collapse and different trajectories of geodesics due to the different geometries.

\begin{figure}
\centering
\includegraphics[scale=0.9]{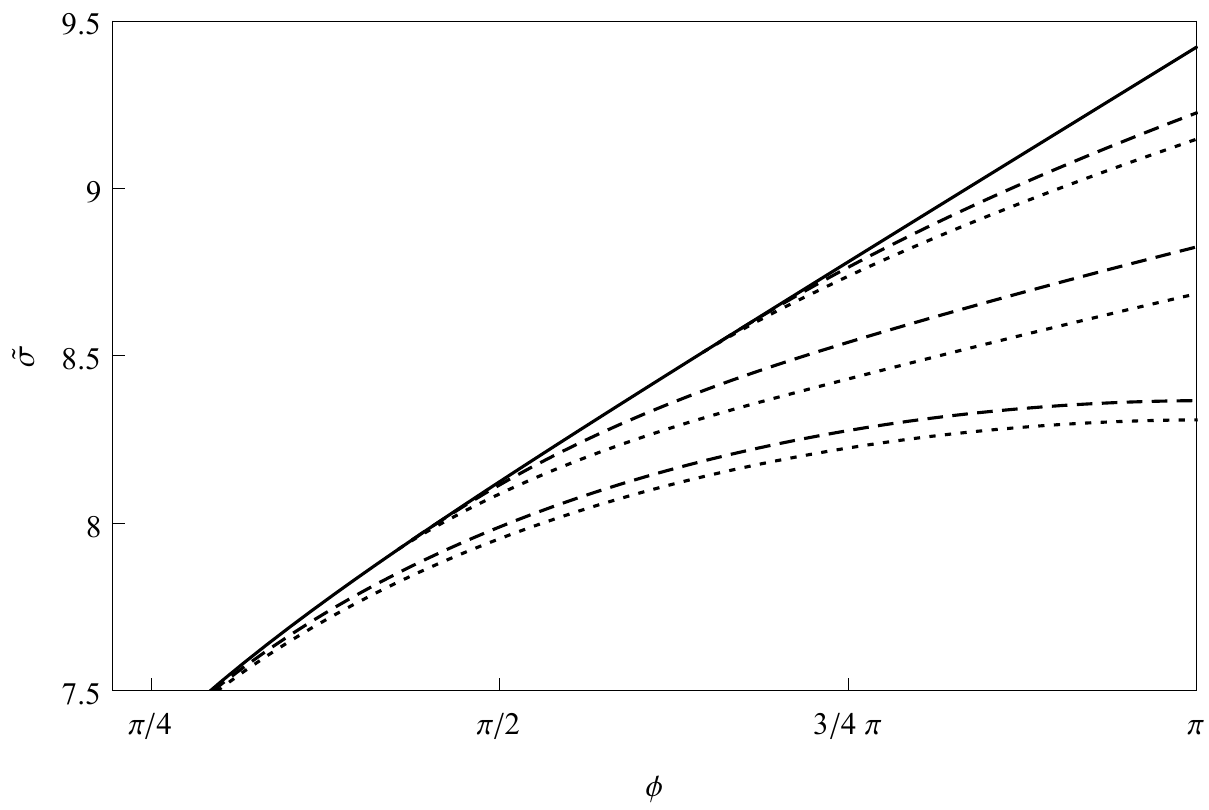}
\caption{Renormalized geodesic length as a function of the angular separation of the boundary points evaluated at different times. Solid line is the equilibrium value, dotted the thin-shell collapse and dashed the Oppenheimer-Snyder collapse. From bottom to upwards the lengths are evaluated at times $t=0$, $0.5$ and $0.75$. The values of other parameters are $n=3$, $m=10$, $r_0=3$ and $\Lambda = 1$.}
\label{fig:multilengths}
\end{figure}

\begin{figure}
\centering
\includegraphics[scale=0.9]{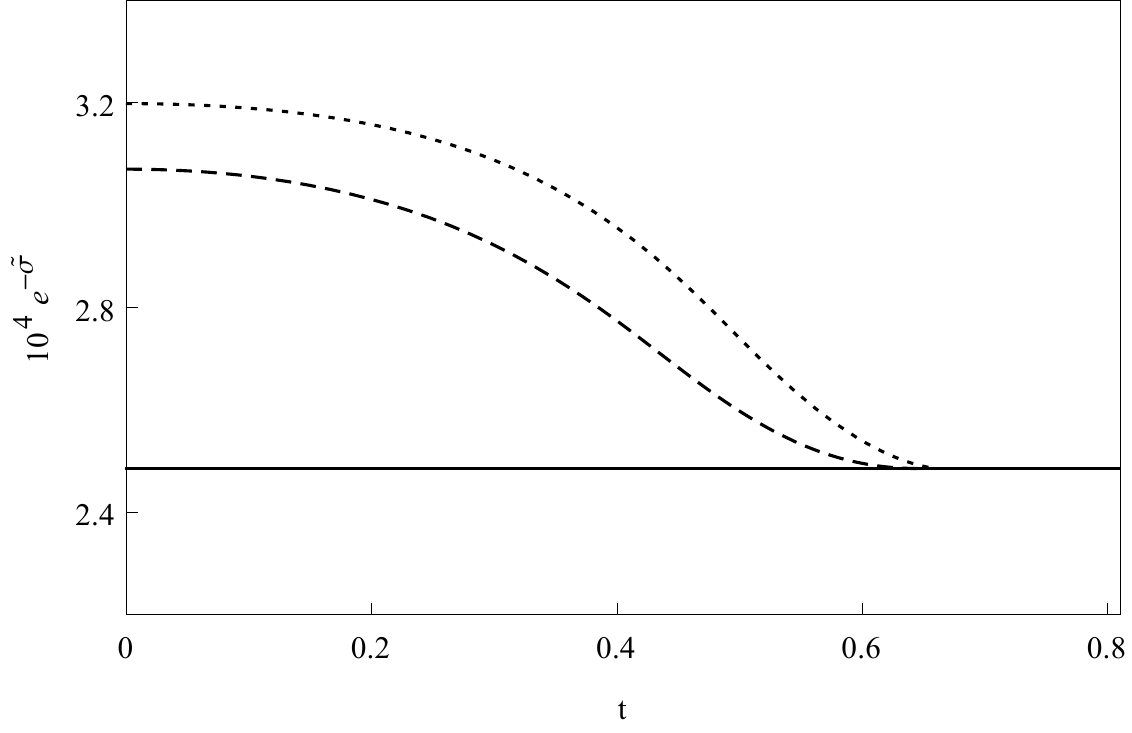}
\caption{The unnormalized correlator $e^{-\tilde{\sigma}}$ for boundary angular separation $\phi = 1.76$ as a function of time. The solid line corresponds to the equilibrium case, the dashed to the Oppenheimer-Snyder case and the dotted to the thin-shell collapse. Other parameters are the same as in figure \ref{fig:multilengths}.}
\label{fig:exp1}
\end{figure}

\begin{figure}
\centering
\includegraphics[scale=0.9]{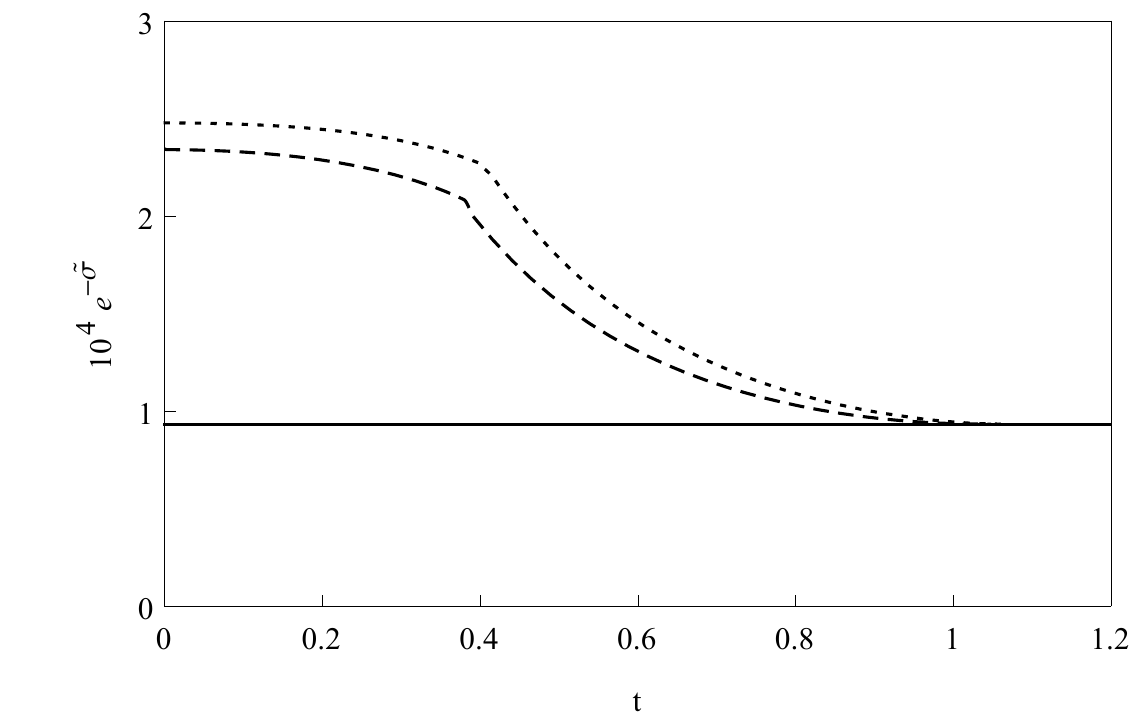}
\caption{The unnormalized correlator $e^{-\tilde{\sigma}}$ for boundary angular separation $\phi = 2.94$ as a function of time. The solid line corresponds to the equilibrium case, the dashed to the Oppenheimer-Snyder case and the dotted to the thin-shell collapse. Other parameters are the same as in figure \ref{fig:multilengths}.}
\label{fig:exp2}
\end{figure}

The motion of the radius of the dust sphere and the shell are compared in figure \ref{fig:trajectories}. Both are released at $t=0$ and start to collapse. The Oppenheimer-Snyder collapse is always faster than the pressureless thin-shell collapse. This can be understood by the distribution of the gravitating matter, since in the OS there is matter in the middle which attracts the outer layers as well. The location of the shell or the edge of the sphere corresponds to the smallest equilibriated scale in the boundary theory, and thus a given scale always thermalizes first in the Oppenheimer-Snyder collapse. It should be noted that the parameters in figure \ref{fig:trajectories} have been chosen to make the difference in the evolution of the two different models to be large, and for generic choices of parameters the actual difference between the two trajectories is even smaller.

Even if the difference between the two different trajectories is small, it does not imply that the difference in the geodesic length between two boundary points is small as well. In the thin-shell model, all the energy-momentum resides at one radial location, $r_s$, which causes a jump in extrinsic curvature normal to the shell. This jump in curvature causes the geodesic to refract as it crosses the shell. In the Oppenheimer-Snyder model there is no delta function in the energy-momentum tensor and correspondingly no jump in extrinsic curvature which would cause the geodesic to refract. Thus the geodesics in the Oppenheimer-Snyder model are more smooth, and geodesics that penetrate the shell deviate less from the equilibrium then in the thin-shell case. It should be noted that due to the coordinate systems in use, this smoothness is not apparent when numerically computing the trajectories of the geodesics.

In figure \ref{fig:multilengths} we have plotted the geodesic length between two points as a function of the angular separation of these two points. The solid line is the equilibrium value, the dashed lines the Oppenheimer-Snyder model and the dotted lines the thin-shell mode. The pairs of dashed and dotted lines correspond to different boundary times. First one can observe that the point where the dashed and dotted line meet the solid line corresponds to the largest scale that has equilibriated, which corresponds to the shell motion in figure \ref{fig:trajectories}. Furthermore, the Oppenheimer-Snyder geodesics are always closer to equilibrium for all angular scales compared to the thin-shell case. Finally, one can note that the difference between the two models is largest in intermediate times, and very early and very late in the gravitational collapse the difference is smaller.

In figures \ref{fig:exp1} and \ref{fig:exp2} we have plotted the exponential of the geodesic lengths of a fixed angular separation as a function of time. According to equation \ref{eq:dictionary}, this is directly proportional to the equal time two-point correlator, modulo an overall normalization factor that has not been determined. Again one can see that a given scale thermalizes faster in the Oppenheimer-Snyder model than in the thin-shell case. It should be noted that the thermalization starts with zero velocity, corresponding to the shell or sphere being initially at rest, but reaches the thermal value in \emph{finite} time of the boundary observer.

\begin{figure}
\centering
\includegraphics[scale=0.7]{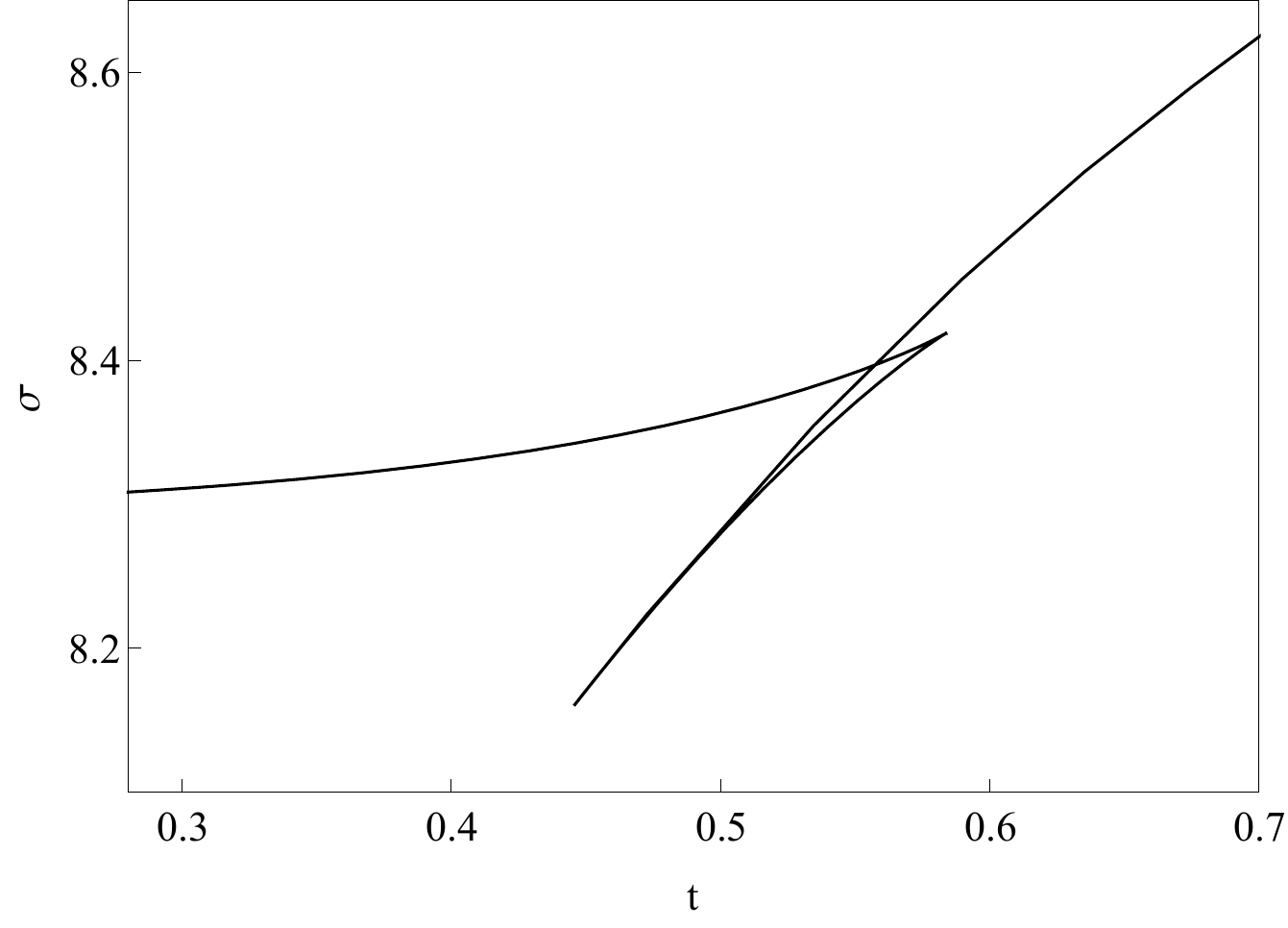}
\caption{The geodesic length between boundary points with angular separation $\phi=2.88$ as a function of time in the OS collapse. For intermediate times there are several geodesics with different lengths connecting the boundary points. The parameters are chosen to be $\Lambda =1$, $m=4$, $n=3$ and $r_0=4$.}
\label{fig:discontinuity}
\end{figure}

When computing the geodesic length, for intermediate times and large angular separation (that is, close to $\varphi = \pi$)  we found the geodesic length to be a multivalued function. That is, it appears that for a given angular separation $\varphi$, there are \emph{several} geodesics that connect these two boundary points. To illustrate this we have plotted the lengths of all geodesics connecting boundary points with the angular separation $\phi =2.88$ in figure \ref{fig:discontinuity}. One can see that for intermediate times there are \emph{three} different lengths corresponding to this fixed angular separation. If the geodesic approximation is taken at face value, the shortest geodesic is the dominant one, and thus the two-point correlator becomes discontinuous. This feature is present in both the Oppenheimer-Snyder and the thin-shell models. The region where the geodesic length seems to be multivalued is always limited to intermediate times and large angles, i.e.~angles close to $\pi$. Similar behaviour was found in the Vaidya model in \cite{Hubeny:2013dea}. Specifially figure \ref{fig:discontinuity} should be compared with figure 7 of \cite{Hubeny:2013dea}. There the authors argue that the geodesic approximation breaks down, and the multivaluedness of $\sigma$ is not to be taken seriously. In the present work we have followed the authors of \cite{Hubeny:2013dea} and assumed that the value of the correlator is given by the geodesic length given by the largest values in figure \ref{fig:discontinuity} and thus it is continuous. Indeed, when we have produced the plots, we have explicitly selected the points corresponding to the continuous values, and discarded the additional points.

Although we can make the correlator thus continuous, its derivative is not: By looking at figure \ref{fig:discontinuity} it is clear that if we choose the two curves that extend to small and large times, their intersection (corresponding to $t \simeq 5.5$ in figure \ref{fig:discontinuity}) has a kink. Thus for small angles where the geodesic lenght is not multivalued, the correlator is smooth as evident in figure \ref{fig:exp1}, but for large angles, it is not, as demonstrated by the kink in figure \ref{fig:exp2}. The kink is present in both the Oppenheimer-Snyder case (dashed line) and the thin-shell case (dotted line).

The discontinuity of the derivative of the correlator has been discussed already in e.g.~\cite{Balasubramanian:2011ur}, where it was argued to be physical. It is interesting to note, however, that both in \cite{Balasubramanian:2011ur} and in \cite{Hubeny:2013dea} the computation was carried out in the Vaidya metric, which has a null hypersurface separating the inside and outside space-times, i.e.~the matter collapses with the speed of light. In this work, however, we have not only looked at time-like shells collapsing, but also the Oppenheimer-Snyder collapse where there is no shell, and thus no discontinuity in the curvature and no delta function inthe energy-momentum tensor. Nevertheless the jump in the time derivative of the correlator is present, demonstrating that the jump does not stem from the null surface in Vaidya nor from the delta-function like distribution of matter. If one assumes that the physical correlator should be smooth, then it seems that the geodesic approximation breaks down not only in the sense that it produces multivaluedness, but also since it does not produce a smooth correlator.

\section{Conclusions}

\label{sec:conclusions}

The Oppenheimer-Snyder model describes the gravitational collapse of a ball of dust and can be used as an analytical model for holographic thermalization. It serves as an alternative to the previously commonly used thin-shell models. One advantage of the OS model is that the OS metric is smoother than the thin-shell model metric due to the lack of a delta-function in the energy-momentum tensor. More important however is that the OS model offers a \emph{different} analytical model for holographic thermalization: As we are interested in the general features of thermalization in holographic models and not features specific to the thin-shell model, the OS model can function as a new comparison to the thin-shell model, and we can hope that the common features of all these models are generic and can be applied to the physical applications of strongly coupled systems.

We investigated the speed of the collapse in the OS model compared to the thin-shell case. We found that the OS model collapses always faster than the corresponding thin-shell model due to the more homogeneous distribution of the energy-momentum. The speed of the collapse is however not an observable, and thus we also computed the equal time two-point function in the geodesic approximation for both cases. With similar initial conditions (same final mass, same initial radius), the OS correlator was found to be at all times closer to equilibrium than the thin-shell correlator. This was due to two separate aspects: First of all the OS model collapses more quickly, but furthermore, since the metric is more smooth at the edge of the sphere than at the location of an infinitely thin-shell, the geodesics are more smooth and refract less, and thus deviate less from the equilibrium geodesics. Thus we concluded that the OS model always thermalizes faster then the thin-shell model.

When we computed the evolution of the radius of the thin shell to compare with the OS model, we  found that for certain non-zero equations of state the thin shell in global AdS has curious trajectories: In addition to the familiar behaviour of a collapse to a black hole or an infinite expansion all the way to the boundary, we found that there are solutions where the shell ends up oscillating between two turning points, thus never collapsing to a black hole but never dissipating to inifinity either. It is unclear what the interpretation of these states are in the boundary field theory, or if these cases correspond to such equations of state which are unattainable in realistic applications of the duality.

When computing the two-point function, we also found that for large times and large angular separations, the geodesic length is multivalued and/or discontinuous. This feature of the geodesic approximation has been previously found in the literature in the Vaidya model \cite{Hubeny:2013dea}, but here we demonstrated for the first time that this discontinuity exists even in the metric which has no delta-functions in the energy-momentum tensor. It seems that the geodesic approximation of the two-point correlator breaks down: not only does the correlator seem multi-valued and discontinuous, but there is also the qualitative issue that the correlator seems to know about physics inside the horizon and thus violating naive understanding of causality. Thus it would be interesting to investigate where exactly the approximation breaks down. If we were investigating the retarded correlator, one could solve a wave equation with certain boundary conditions to obtain the full correlator and then see how this compares with the approximation. However, for the correlator at hand such a simple calculation is not possible. Instead, a procedure described in e.g.~\cite{Keranen:2014lna} is required, which given the non-trivial metric seems a very difficult task.

The main interest in this and previous work is to discover which features of the analytical models are generic enough that they can be expected to be present in physically realistic scenarios (e.g.~heavy ion collisions). For the equal time two-point correlator we found out that results are qualitatively very similar in both OS and thin-shell models. We also found that the apparent breakdown of the geodesic approximation is present in both models. The next step would be to compute the different observables for the OS model which have already been computed in the thin-shell model in the literature \cite{Lin:2008rw,Baier:2012ax,Baier:2012tc} and see which features of the thermalization are in fact just features of the thin-shell model. For example the photon production rate and dilepton spectrum computed in \cite{Baier:2012ax,Baier:2012tc} exhibited oscillations which were speculated to be related to the presence of the delta-function-like shell. Performing the same computation in the OS model would immediately determine the source of the oscillatory behaviour.

\acknowledgments

The author would like to thank Ville Ker\"{a}nen and Javier Mas Sol\'{e} for useful discussions, and Aleksi Vuorinen for comments on the manuscript.
This work is part of the research program of the "Stichting voor Fundamenteel Onderzoek der Materie (FOM)", which is financially supported by the "Nederlandse organisatie voor Wetenschappelijke Onderzoek (NWO)". The author would also like to thank the hospitality of the Helsinki Institute of Physics.

\appendix

\section{The continuous coordinate system}

\label{sec:continuous}

To compute the matching conditions, e.g.~how a geodesic travers the shell from one coordinate system to another, we construct a coordinate system which is explicitly continuous at the edge of the sphere. This coordinate system uses coordinates $\tau$ and $\lambda$ as its time and radial coordinates. $\lambda$ is defined as the proper distance from the hypersurface which is the edge of the sphere \emph{normal to the hypersurface}. In more practical terms, it is defined as the length of the geodesic that at the hypersurface is normal to the hypersurface, i.e.~$dx^\mu/d\lambda = n^\mu$. The time coordinate $\tau$ is then defined to be the proper time of the shell when that that geodesic intercepts the hypersurface. Although this coordinate system has a very complicated form in the bulk, at the hypersurface joining the inside and outside metric it takes the form of a Minkowski metric. We can then transform the geodesic both inside and outside to this coordinate system and propagate the geodesic infinitesimally across the hypersurface. Since the metric is continuous, the geodesic evolves smoothly. The only ingredients we then need for this computation are the partial derivatives of the different coordinates, which we can use to perform this transformation. For the outside (and inside in the thin shell case) metric these are given by
\begin{equation}
\PD{r,\tau,\lambda} = \dot{r}_s\;,\;
\PD{r,\lambda,\tau} = \sqrt{f+\dot{r}_s}\;,\;
\PD{t,\tau,\lambda} = \frac{\sqrt{f+\dot{r}_s^2}}{f}\;\;\text{and}\;
\PD{t,\lambda,\tau} = \frac{\dot{r}_s}{f}\;.
\end{equation}
For the FRW metric inside the sphere in the Oppenheimer-Snyder case, the derivatives are given by
\begin{equation}
\PD{a,\tau,\lambda} = -\sqrt{h(a)}\;,\;
\PD{\rho,\tau,\lambda} = 0\;,\;
\PD{a,\lambda,\tau} = 0\;\;\text{and}\;
\PD{\rho,\lambda,\tau} = \frac{\sqrt{1+\rho^2}}{a}\;.
\end{equation}

\section{Eddington-Finkelstein coordinate system}

\label{sec:eddingtonfinkelstein}

When computing the geodesics for the equal time two-point correlators, one needs to solve for geodesics that cross the shell or leave the sphere only after the horizon has been formed. Easiest way to avoid the divergences associated with the singularity at the horizon is to use the Eddington-Finkelstein metric for the outside metric, given by
\begin{equation}
ds^2 = -f(r) \, dv^2 +2\,dv\,dr +r^2\,d\Omega^2_n \; .
\end{equation}
The geodesic equations for a space-like geodesic in this coordinate system can be integrated to be
\begin{align}
r^2 \dot{\phi} & = L\\
\dot{r}^2 & = f(r)\left[ 1-\frac{L^2}{r^2}\right] + E^2\\
\dot{v} & = \frac{E+\dot{r}}{f(r)} \; .
\end{align}
The location of the shell $(v_s(\tau),r_s(\tau),\ldots)$ obeys
\begin{equation}
\dot{r}_s = \frac{1}{2}\left(f\,\dot{v}_s-\frac{1}{\dot{v}_s} \right) \qquad \text{and} \qquad \dot{v}_s = \frac{\dot{r}_s + \sqrt{f+\dot{r}_s^2}}{f} \; .
\end{equation}
The normal vector of the shell is
\begin{equation}
n^r = \dot{v}_s\left( f- \frac{\dot{r}_s}{\dot{v}_s}\right)\qquad \text{and} \qquad n^v = \dot{v}_s \; .
\end{equation}
The coordinate transformation to the continuous coordinate system $(\tau,\lambda)$ at the shell or the edge of the sphere is then given by
\begin{align}
\PD{r,\lambda,\tau} & = \sqrt{f+\dot{r}_s^2} &
\PD{v,\lambda,\tau} &= \frac{\dot{r}_s+\sqrt{f+\dot{r}_s^2}}{f}\\
\PD{r,\tau,\lambda} & = \dot{r}_s&
\PD{v,\tau,\lambda} & = \frac{\dot{r}_s+\sqrt{f+\dot{r}_s^2}}{f}
\end{align}

\end{document}